\documentclass[superscriptaddress,twocolumn,english,floatfix,aps,prd,amsmath,amssymb,longbibliography]{revtex4-2}
\usepackage[T1]{fontenc}
\usepackage[latin9]{inputenc}
\usepackage{babel}
\usepackage{polski}
\usepackage{color}
\usepackage{times}
\usepackage{epsfig}
\usepackage{subfigure}
\usepackage{amsmath}
\usepackage{amssymb}
\usepackage{graphicx}
\usepackage[colorlinks=true]{hyperref}
\hypersetup{citecolor=blue,linkcolor=blue,urlcolor=blue}

\begin{document}
\title{Vacuum energy around particles}

\author{Danilo T. Alves}
\email{danilo@ufpa.br}
\affiliation{Faculdade de F\'{i}sica, Universidade Federal do Par\'{a}, 66075-110, Bel\'{e}m,
Brazil}
\date{\today}
\begin{abstract}
The creation of particles by the excitation of the quantum vacuum in a cavity with a moving mirror was predicted in 1969.
Here, we investigate that, in addition to real particles, the excitation of the quantum vacuum in a dynamical cavity can also result in the creation of a certain amount of positive vacuum energy around these particles. 
We show that while in the case of a single mirror moving in a free space, with the field in the vacuum state, all the energy transferred from the mirror to the field is converted into real particles, 
on the other hand, when considering a second and static mirror forming a cavity with the first, the same movement of the first mirror can lead to less energy being converted into real particles, with the difference being converted into positive vacuum energy around these particles.
\end{abstract}
\maketitle

\section{Introduction}
\label{sec:intro}
Considering a real massless scalar field in a two-dimensional spacetime,
Moore \cite{Moore-1970} predicted in 1969 the creation of particles by the excitation of the quantum vacuum in a cavity with a moving mirror. 
This effect, usually known as the dynamical Casimir effect (DCE),
was investigated in other pioneering articles (for example, in Refs. \cite{DeWitt-PhysRep-1975,Fulling-Davies-PRSA-1976,Davies-Fulling-PRSA-1977,Davies-Fulling-PRSA-1977-II,Candelas-JMP-1976,Candelas-PRSA-1977}) and,
since then, hundreds of articles have been written on this issue
(see Refs. \cite{Dalvit-CasimirPhysics-2011,Dodonov-Phys-2020} and those therein), with some experiments already done \cite{Wilson-Nature-2011,Lahteenmaki-2013, Vezzoli-2019,Schneider-et-al-PRL-2020}.
Beyond the DCE, there are other effects of particle creation from the quantum vacuum in non-stationary systems as, for example, in models of expanding universe and rotating black holes \cite{Davies-JOPTB-2005}.

\begin{figure}
	\centering
	\subfigure[\label{fig:visual-livre}]{\epsfig{file=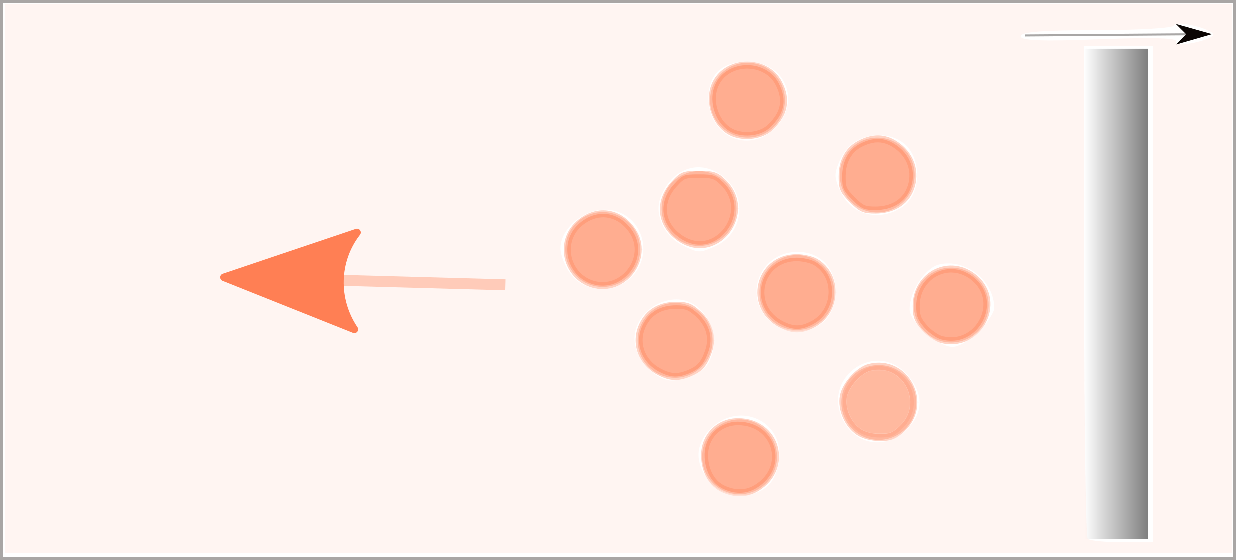, width=0.48 \linewidth}}
	\subfigure[\label{fig:visual-cavidade}]{\epsfig{file=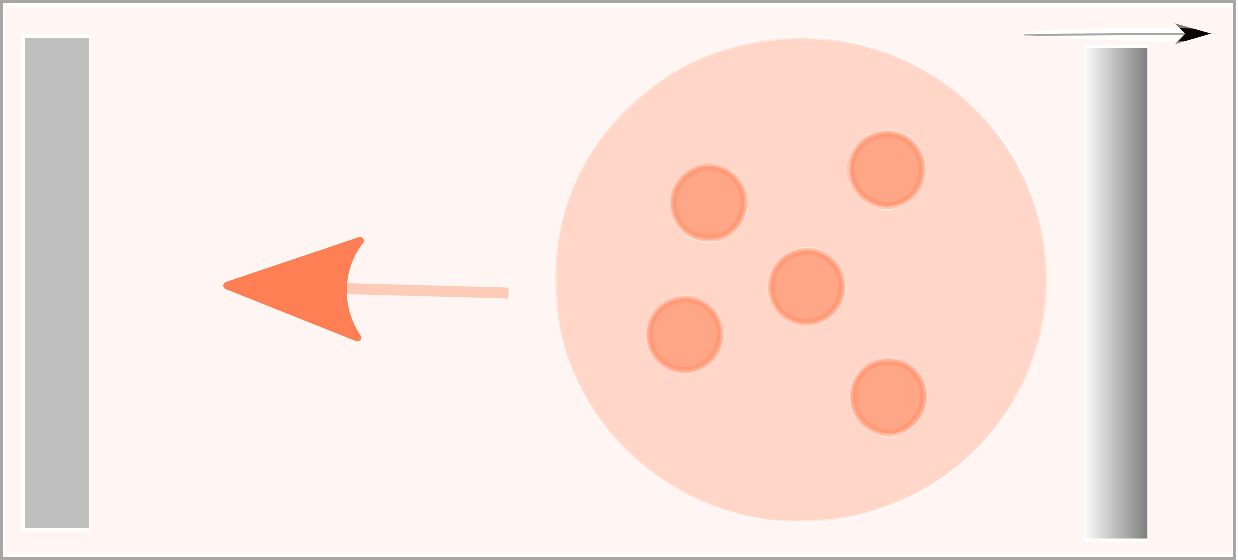, width=0.48 \linewidth}}
	\caption{
		In (a), we illustrate the case of a free space, 
		where all the energy dissipated from the moving mirror is converted into real particles (small circles).
		In (b), we illustrate the case of a cavity with the same movement of the moving mirror and the same amount of
		dissipated energy as in (a), but with only a part of this energy converted into real particles (smaller number of small circles), with the difference being converted into positive vacuum energy (big circle) propagating together with these particles.
	}
	\label{fig:visual}
\end{figure}

In the present paper, we investigate that, in addition to real particles, the excitation of the quantum vacuum in a dynamical cavity can also produce an amount (a ``cloud'') of positive vacuum energy around these particles. 
Considering the same model of
a real massless scalar field in a two-dimensional spacetime
used by Moore \cite{Moore-1970}, we show that when one has only one mirror moving in a free space, with the field in the vacuum state, all the energy dissipated from the mirror is converted into real particles [Fig. \ref{fig:visual-livre}], but, when considering a second and static mirror, the same movement of the first mirror, with the same dissipated energy,
can lead to less energy being converted into real particles, with the difference being converted into positive vacuum energy around the particles [Fig. \ref{fig:visual-cavidade}].
The basis of our investigation is an expanding one-dimensional cavity, in which one of the mirrors, initially at rest, starts moving smoothly, acquires constant velocity for a certain time interval, after which it smoothly stops, returning to rest.
We focus on this configuration because one can define in- and out-states, separate the vacuum regions from those where the particles can be present \cite{DeWitt-PhysRep-1975,Fulling-Davies-PRSA-1976,Castagnino-Ferraro-AnnPhys-1984},
and the field energy density is well defined for all spacetime points.

\section{Basic ideas}
\label{sec:fundamental-formulas}
We start by reviewing some fundamental formulas and techniques of calculation found in Refs. \cite{Moore-1970, Fulling-Davies-PRSA-1976, Cole-Schieve-PRA-1995, Cole-Schieve-PRA-2001, Alves-Granhen-Silva-Lima-PRD-2010, Alves-Granhen-CPC-2014}.
We also present original discussions, useful for the purposes of the present paper.
We consider a real massless scalar field in a cavity, satisfying the wave equation
$
\left(\partial _{t}^{2}-\partial _{x}^{2}\right) \phi \left(
t,x\right) =0
$
(we assume throughout this paper $\hbar=c=1$),
and obeying the Dirichlet boundary condition imposed at a static mirror located at $x=0$,
and also at the moving mirror's position at $x=L(t)$ (i.e.
$\phi\left( t,0\right)=0$ and $\phi[t,L(t)]=0$),
where $L(t)>0$ is an arbitrary prescribed law
for the moving mirror (hereafter named right mirror), with $L(t<0)=L_0$,
where $L_0$ is the length of the cavity in the initial ($t<0$) static situation 
(in-state).
The exact formula for the expected value ${\cal T}(t,x)$ of the energy 
density operator $\hat{T}_{00}(t,x)$, i.e. ${\cal T}(t,x)=\langle\hat{T}_{00}(t,x)\rangle_{\text{in}}$
(the averages $\langle...\rangle_{\text{in}}$ are taken over a vacuum in-state in the cavity,
which is our interest in this paper), can be written recursively by
the equations (see Refs. \cite{Cole-Schieve-PRA-1995,Cole-Schieve-PRA-2001, Alves-Granhen-Silva-Lima-PRD-2010,Alves-Granhen-CPC-2014} and Appendix \ref{ap:basic-formulas}):
\begin{gather}
	{\cal T}(t,x)={\cal{T}}_{{\cal \tilde{A}}}\left(t,x\right)+
	{\cal{T}}_{{\cal \tilde{B}}}\left(t,x\right),
	\label{eq:T}\\
	{\cal{T}}_{{\cal \tilde{A}}}\left(t,x\right)=-{\pi}/({48L_{0}^{2}})[{\cal \tilde{A}}(u)+{\cal \tilde{A}}(v)],\label{eq:T-A-tilde}\\ 
	{\cal{T}}_{{\cal \tilde{B}}}\left(t,x\right)=-{\cal \tilde{B}}(u)-{\cal \tilde{B}}(v),\\
	{\cal \tilde{A}}(z)=\prod_{i=1}^{n(z)}{\cal A}[t_{i}(z)],\label{eq:A-tilde}\\ 
	{\cal \tilde{B}}(z)=\sum_{j=1}^{n(z)}{\cal B}[t_{j}(z)]\prod_{i=1}^{j-1}{\cal A}[t_{i}(z)],\label{eq:B-tilde}\\
	%
	{\cal{A}}(t)=\left[ \frac{1-\dot{L}(t)}{1+\dot{L}(t)}\right]^{2}\label{eq:f-A},\\
	%
	{\cal B}(t)=\frac{-\frac{\dddot{L}\left(t\right)}{12\pi}-\frac{1}{4\pi}\frac{\ddot{L}^{2}\left(t\right)\dot{L}\left(t\right)}{\left[1-\dot{L}\left(t\right)^{2}\right]}}{\left[1+\dot{L}\left(t\right)\right]^{3}\left[1-\dot{L}\left(t\right)\right]},
	\label{eq:f-B}
\end{gather}
where $u=t-x$, $v=t+x$, $n(z)$ is the number of reflections off the moving mirror, necessary to connect a null line $v=z$ (or $u=z$) to a null line in the static zone [
$v<L_0$, where $f(v)={\pi}/({{48}L_0^2})$], and $t_1, ..., t_n$ are such that
the points $[t_i,L(t_i)]$ are those where the null lines $v=t_i+L(t_i)$ intersect the right mirror, and
are calculated via $z=t_1+L(t_1)$, and $t_{i+1}+L(t_{i+1})=t_i-L(t_i),\;i=1,2,3,...$
(for more details, see Refs. \cite{Cole-Schieve-PRA-1995,Cole-Schieve-PRA-2001, Alves-Granhen-Silva-Lima-PRD-2010,Alves-Granhen-CPC-2014}).

When the right mirror stops its motion at an instant $t=\beta$, with the cavity reaching a final length $L_{\text{fin}}$ (out-state), we have the total energy $E$ given by $E=E_{\cal \tilde{A}}+E_{\cal \tilde{B}}$, with
$E_{\cal \tilde{A}}$ $=$ $\int_{0}^{L_{\text{fin}}}{\cal{T}}_{{\cal \tilde{A}}}\left(t,x\right) dx$, and
$E_{\cal \tilde{B}}=\int_{0}^{L_{\text{fin}}}{\cal{T}}_{{\cal \tilde{B}}}\left(t,x\right)dx$.
On the other hand, $E$ can also be written as (see Ref. \cite{Alves-Granhen-Silva-Lima-PRD-2010} and Appendix \ref{ap:basic-formulas}) 
$E=E^{\text{(cas)}}+E^{\text{(par)}}$,
where $E^{\text{(cas)}}$ is the Casimir energy in the out-state (out-Casimir energy), 
given by $E^{\text{(cas)}}=-{\pi}/({24L_{\text{fin}}})$
[hereafter, the superscripts ``$\text{cas}$'' and ``$\text{cas-in}$''
mean Casimir energy in out- and in-states, respectively], 
and the total energy of real particles $E^{\text{(par)}}$ is given by $E^{\text{(par)}}=\sum_{n}\omega_{n}{N}_n$,
with $\omega_{n}=n\pi/L_{\text{fin}}$, and  ${N}_n=\left\langle \hat{a}_{n}^{\dag }\hat{a}_{n}\right\rangle_{\text{in}}$ is the mean number of particles related to the $n$th field mode in a cavity with length $L_{\text{fin}}$, $\hat{a}_{n}^{\dag}$ and $\hat{a}_{n}$ are, respectively, the creation and annihilation
operators related to the $n$th field mode.
We highlight that, independently of the field state, the total vacuum energy in the static cavity is the Casimir one. Using both formulas for $E$, we have
\begin{gather}
	E_{\cal \tilde{A}}+E_{\cal \tilde{B}}
	=E^{\text{(cas)}}+E^{\text{(par)}}.
	\label{eq:AA-BB-EE}
\end{gather}
%
%
This formula is fundamental to the discussion
presented here.

\section{Missing Casimir energy and positive vacuum energy around particles}
\label{sec:missing-and-positive}

\begin{figure}
	\centering
	\subfigure[\label{fig:FD}]{\epsfig{file=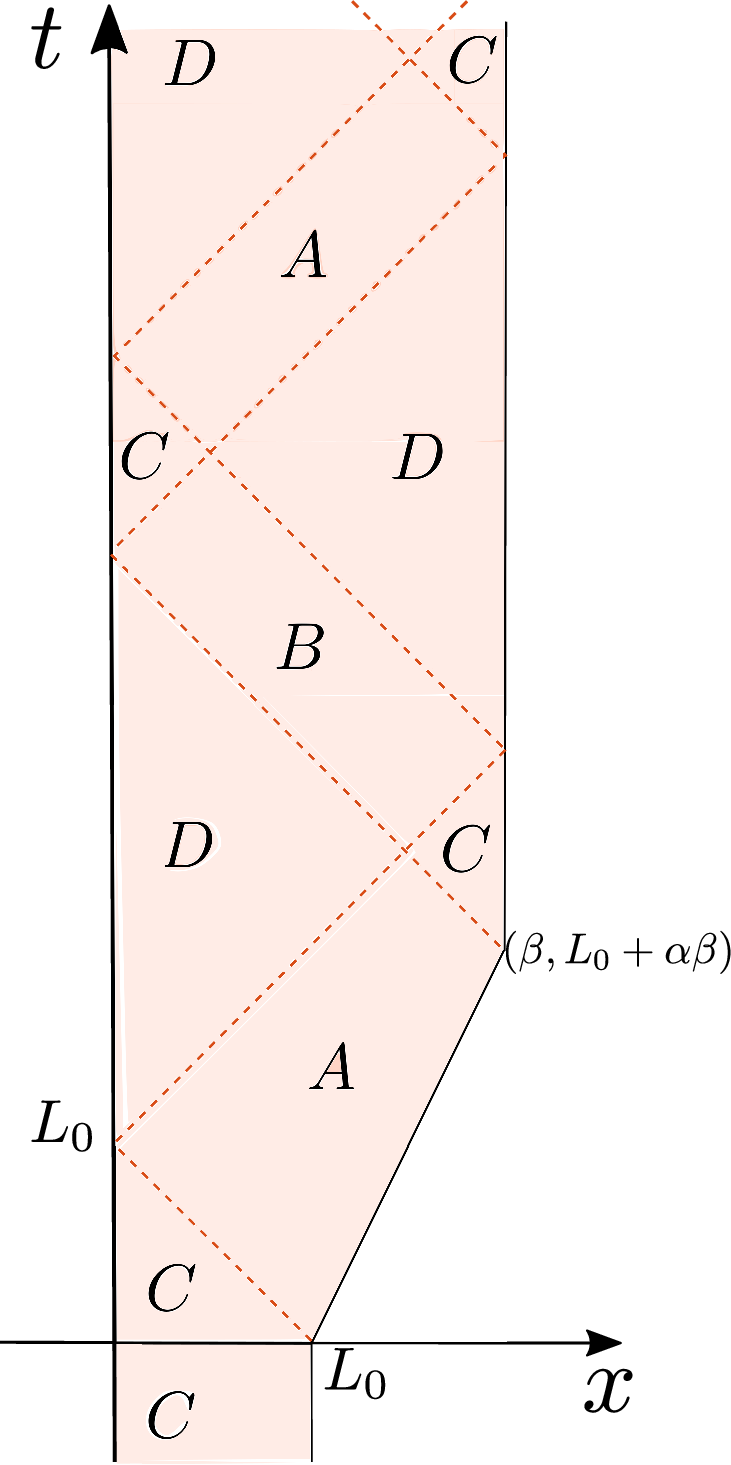, width=0.39 \linewidth}}
	\subfigure[\label{fig:FD-suave}]{\epsfig{file=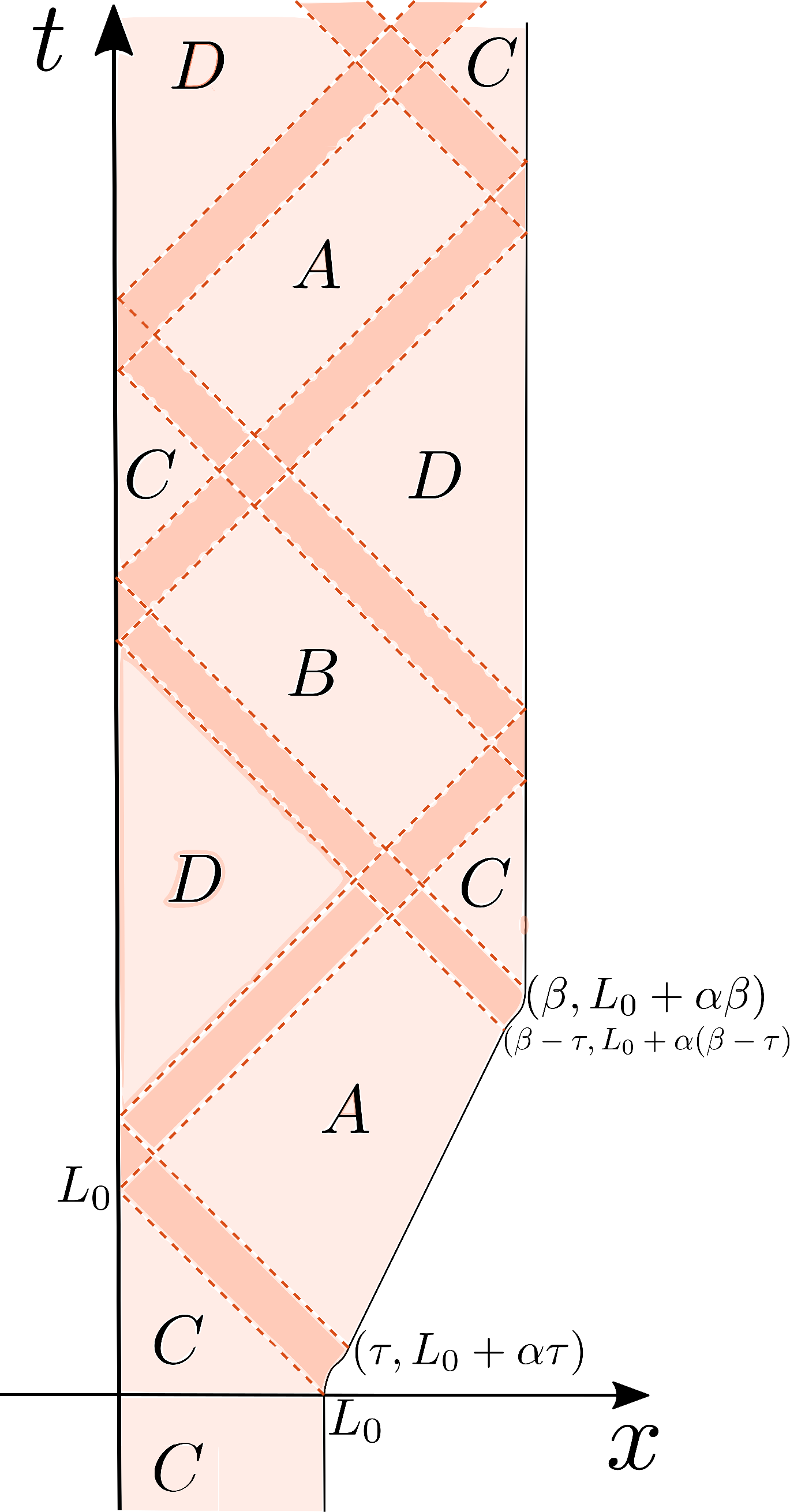, width=0.41 \linewidth}}
	\caption{
		In (a) and (b): the solid lines are worldlines of two mirrors, with one of them remaining static at $x=0$, and the other following the law of motion given in Eq. \eqref{eq:FD-suave}, with $\tau=0$ (a) and $\tau\neq0$ (b)
		(figure (a) was adapted from that found in Ref. \cite{Fulling-Davies-PRSA-1976}); the lighter regions $A$, $B$, $C$ and $D$ are vacuum regions.
		In (a), the dashed lines (particle regions) are the worldlines (null lines) of the massless particles emitted from the abrupt accelerations of the right	mirror.
		In (b), the darker regions (particle regions) are regions where the created particles can be found. 
}
	\label{fig:FD-e-FD-suave}
\end{figure}

Fulling and Davies \cite{Fulling-Davies-PRSA-1976} investigated the model of a real massless scalar field in 1+1 dimensions in a cavity in which one of the mirrors remains at rest at $x=0$, and the other starts moving abruptly at $t=0$, 
acquiring a constant velocity $\alpha$ $(0<\alpha<1)$, and stopping abruptly at $t=\beta$. The vacuum was considered the initial field state $(t<0)$.
The worldlines of the mirrors are shown,
for $\beta<2L_0/\left(1-\alpha\right)$, in Fig. \ref{fig:FD}
(which was adapted from Fig. 4(b) found in Ref. \cite{Fulling-Davies-PRSA-1976}).
The dashed lines, which have slopes $1$ or $-1$, are the worldlines (null lines) of the massless particles emitted 
from the abrupt accelerations of the right mirror.
These idealized accelerations, proportional to Dirac $\delta$ functions, generate an undefined value of the field energy density in these lines, and it also can be shown that this motion produces a divergent 
expected value of the total number of created particles \cite{Castagnino-Ferraro-AnnPhys-1984}.
On the other hand, regions $A$, $B$, $C$ and $D$, separated by the dashed lines
[Fig. \ref{fig:FD}], are vacuum regions
(since there is no particle production during the uniform motion of the mirror \cite{DeWitt-PhysRep-1975,Fulling-Davies-PRSA-1976,Castagnino-Ferraro-AnnPhys-1984})
with well defined field energy densities given by \cite{Fulling-Davies-PRSA-1976}:
${\cal T}_{A}={\cal T}_{B}=-\left[\pi/(24 L_0^2)\right]\left[\left(1+\alpha^{2}\right)/\left(1+\alpha\right)^{2}\right]$,
${\cal T}_{C}$$={\cal T}^{\text{(cas-in)}}=-\pi/(24 L_0^2)$,
${\cal T}_{D}=$$-\left[\pi/(24 L_0^2)\right]\left[\left(1-\alpha\right)/\left(1+\alpha\right)\right]^{2}$,
where ${\cal T}^{\text{(cas-in)}}$ is the Casimir field energy density
in the in-state. These equations can be recovered using Eqs. \eqref{eq:T}-\eqref{eq:f-B}
(see Appendix \ref{ap:recovering-FD}). 
%

In order to work with a well defined
${{\cal T}}\left(t,x\right)$ $\forall (t,x)$,
we will discuss models of expanding cavities which do not generate the mentioned divergences occurring at $t=0$ and $t=\beta$.
Thus, we propose laws of motion that smooth the jumps in the derivatives of $L(t)$
occurring at these instants, so that
$\dot{L}\left(t\right)$, $\ddot{L}\left(t\right)$,
and $\dddot{L}\left(t\right)$ are well defined functions $\forall\;t$.
In this way, the energy density given in Eq. \eqref{eq:T}
is well defined in all spacetime points, so that singularities are no more present in the regions where the created particles can be.
Trajectories $x=L(t)$ that do not lead to singularities are, for example, given by (see Appendix \ref{ap:smoothing})
\begin{equation}
	L\left(t\right)=\begin{cases}
		L_0 & \left(t<0\right),\\
		L_{1}\left(t\right) & \left(0\leq t \leq\tau\right),\\
		L_0+\alpha t & \left(\tau<t<\beta-\tau\right),\\
		L_{2}\left(t\right) & \left(\beta-\tau\leq t\leq \beta\right),\\
		L_0+\alpha\beta & \left(\beta<t\right),
	\end{cases}
	\label{eq:FD-suave}
\end{equation}
where
$L_{1}\left(t\right)=\;e^{t}-1+L_0-t-\frac{1}{2}t^{2}-\frac{1}{6}t^{3}
+a_{1}t^{4}+b_{1}t^{5}+c_{1}t^{6}+d_{1}t^{7}$, and
$L_{2}\left(t\right)=e^{\left(t-\beta\right)}-1+L_0+\alpha\beta-\left(t-\beta\right)-\frac{1}{2}\left(t-\beta\right)^{2}
-\frac{1}{6}\left(t-\beta\right)^{3}+a_{2}\left(t-\beta\right)^{4}+b_{2}\left(t-\beta\right)^{5}
+c_{2}\left(t-\beta\right)^{6}+d_{2}\left(t-\beta\right)^{7}$.
Note that, for $\tau=0$, we recover
the mentioned law of motion investigated by Fulling and Davies \cite{Fulling-Davies-PRSA-1976}.

Hereafter, our discussions assume (but are not restricted to) functions $L(t)$ according to the 
law of motion in Eq. \eqref{eq:FD-suave}, which is illustrated
in Fig. \ref{fig:FD-suave}. We assume that $\beta<2L_{0}/\left(1-\alpha\right)$, which means that particles can be created
\cite{Fulling-Davies-PRSA-1976,Castagnino-Ferraro-AnnPhys-1984} (see also Appendix \ref{ap:recovering-FD}), and that a field mode, after being affected by the moving mirror, never meets this mirror in motion again, which allows us to exclude the effect of reinteraction with the moving mirror on the particle creation process \cite{Alves-Granhen-Alves-Lima-PRD-2020}.

For any instant $t>\beta$, the cavity is static again, with length $L_{\text{fin}}=L_0+\alpha\beta$, and the field is in an out-state different from vacuum.
One can see that $C$ regions [Figs. \ref{fig:FD} and \ref{fig:FD-suave}] are vacuum, but with an energy density smaller than the out-Casimir one: ${\cal T}_{C}<{\cal T}^{\text{(cas)}}$, where ${\cal T}^{\text{(cas)}}=-\pi/(24L_{\text{fin}}^2)$. 
On the other hand, ${\cal T}_{D}>{\cal T}^{\text{(cas)}}$, so that $D$ regions are vacuum, 
but with an energy density greater than the out-Casimir one.
Regions $A$ and $B$ are also vacuum, where ${\cal T}_{A}$ and  ${\cal T}_{B}$ can be smaller, equal or greater than ${\cal T}^{\text{(cas)}}$, 
depending on the parameters $\alpha$ and $\beta$. 
It is also interesting to highlight that in the out-state the $C$ regions
can be viewed as a ``fossil'', or an
explicit signature, of the in-state of the field in the cavity.

Before we go on, let us define some conventions to be used hereafter.
The superscript ``$\text{vac}$'' indicates 
that the energy (or density) comes from vacuum (in out-state).
This superscript, combined with the subscripts ``$\text{vac}$'' and 
``$\text{par}$'', indicates that this vacuum energy (or density) is located in vacuum or particle regions (those where real particles can be found), respectively.
The superscript ``$\text{par}$'' indicates 
that the energy (or density) comes from the real particles (out-state),
which is just different from zero in the particle regions.
$\overline{{\cal T}}$
means the average (in space) of a given energy density, in a given region of the cavity.

\begin{figure}
	{\epsfig{file=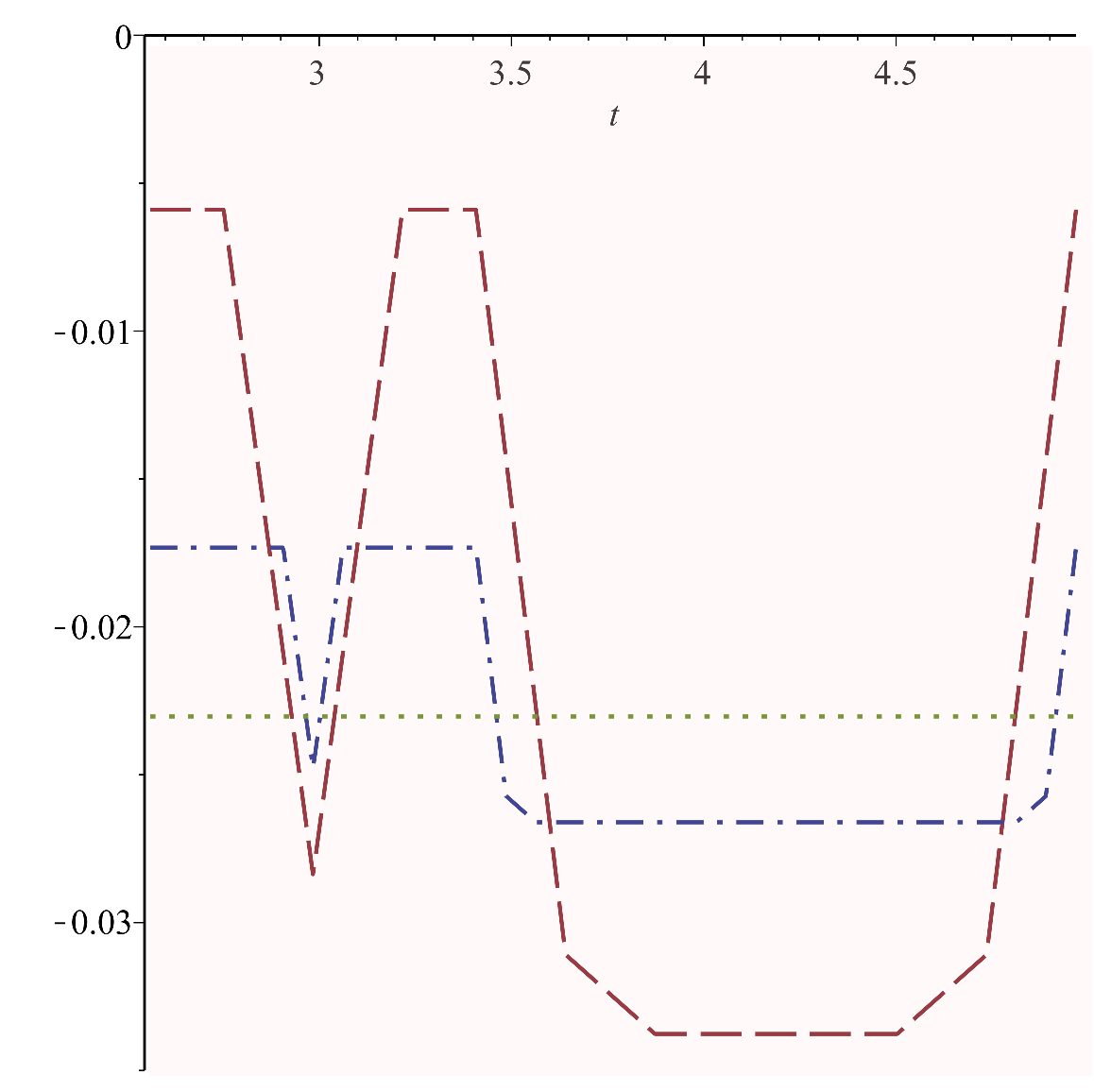, width=0.50 \linewidth}}
	\caption{
	Missing Casimir energy $\Delta E_{\text{(vac)}}^{\text{(vac)}}(t,\tau)$
	(vertical axes) versus $t$ (horizontal axes), for $L_0=1$, $\alpha=0.55$, and $\beta=2.56$.
	The dotted, dash-dotted, and  dashed  lines show
	the cases $\tau=0$,  $\tau=10^{-1}$, and $\tau=3\times10^{-1}$, respectively. 
}
\label{fig:Delta-E}
\end{figure}

Now, let us define the difference between the vacuum energy in the vacuum
regions, and the corresponding out-Casimir energy in these regions (both in the out-state):
$\Delta E_{\text{(vac)}}^{\text{(vac)}}(t,\tau)=E_{\text{(vac)}}^{\text{(vac)}}(t,\tau)-E_{\text{(vac)}}^{\text{(cas)}}(t,\tau)$.
For $\tau=0$ [Fig. \ref{fig:FD}], we have
$\Delta E_{\text{(vac)}}^{\text{(vac)}}(t,0)=
-{\pi}/{(24L_{0}^{2})}{[\alpha^{2}\beta\left(\alpha\beta+2{ L_{0}}-\beta\right)]}/[\left(1+\alpha\right)\left(\alpha\beta+{L_{0}}\right)]$,
with $-\pi/(24L_0)<$ $\Delta E_{\text{(vac)}}^{\text{(vac)}}(t,0)$ $<0$
(see the dotted line in Fig. \ref{fig:Delta-E}).
Since $E^{\text{(cas)}}_{\text{(par)}}(t,0)=0$ and, as already discussed, the total vacuum energy is the out-Casimir one, this implies that an amount of positive vacuum energy $-\Delta E_{\text{(vac)}}^{\text{(vac)}}(t,0)$
must be in  the particle regions [the null lines in Fig. \ref{fig:FD}], so that
$E_{\text{(par)}}^{\text{(vac)}}(t,0)=-\Delta E_{\text{(vac)}}^{\text{(vac)}}(t,0)$.
The amount $\Delta E_{\text{(vac)}}^{\text{(vac)}}(t,0)$, we call ``missing Casimir energy''
in the vacuum regions. 
Defining $x_a(t)$ and $x_b(t)$ the $x$ coordinates of the particle regions, we can write $\overline{{\cal T}}_{\text{(par)}}^{\text{(vac)}}\left(t,x\right)=-\frac{1}{2}\Delta E_{\text{(vac)}}^{\text{(vac)}}\left\{ \delta\left[x-x_{a}\left(t\right)\right]+\delta\left[x-x_{b}\left(t\right)\right]\right\}$.

\begin{figure}
	\centering
	\subfigure[\label{fig:T-3-10-1}]{\epsfig{file=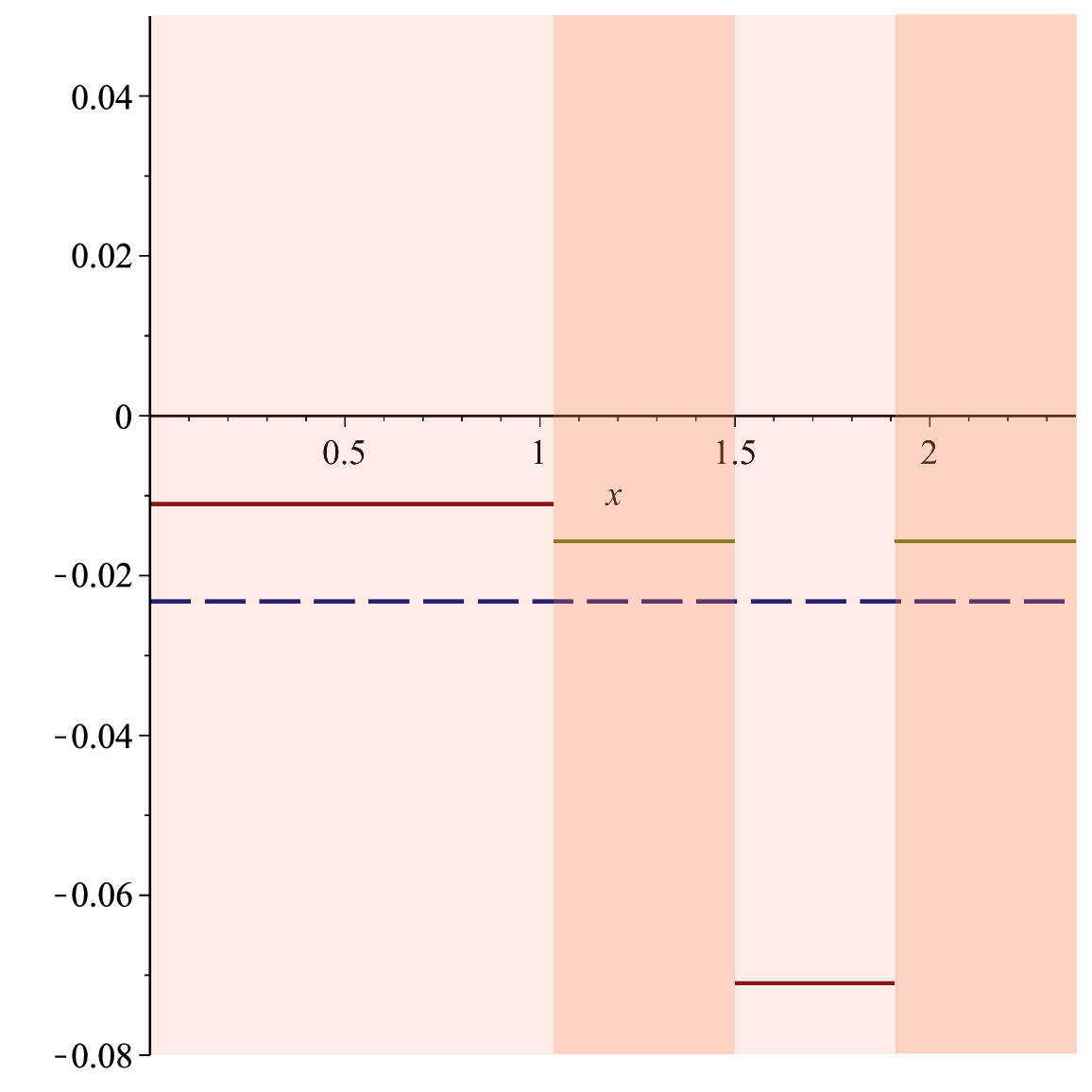, width=0.49 \linewidth}}
	\subfigure[\label{fig:T-10-1}]{\epsfig{file=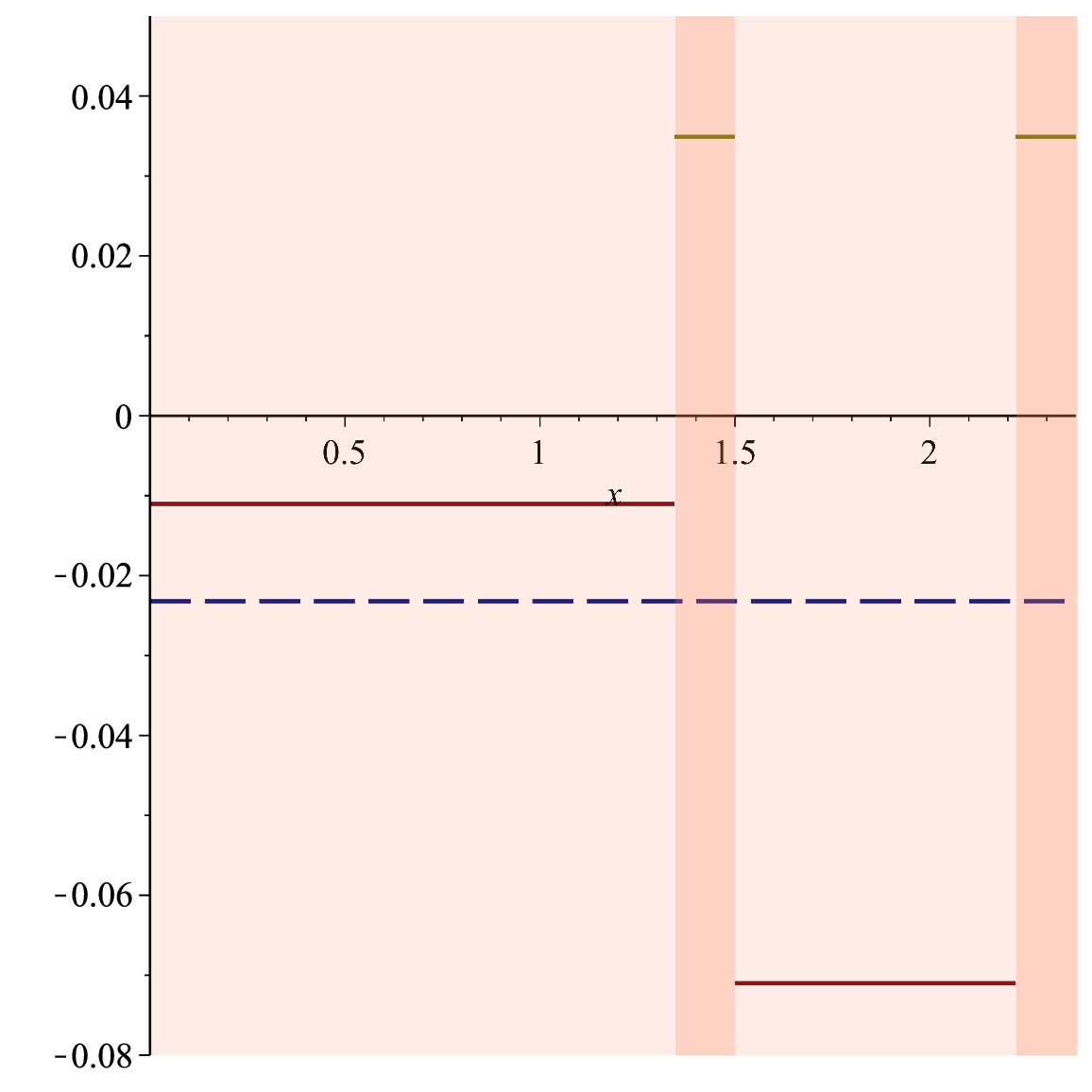, width=0.49 \linewidth}}
	\caption{
	Energy densities (vertical axes) at $t=\beta$, versus $x$ (horizontal axes), for
	$0\leq x\leq L_0+\alpha\beta$, considering $L_0=1$, $\alpha=0.55$, and $\beta=2.56$.
	In both figures lighter regions are vacuum regions, whereas darker regions are particle regions.
	The solid red lines in lighter regions show ${\cal T}_{\text{(vac)}}^{\text{(vac)}}\left(\beta,x\right)$,
	the solid green lines in darker regions show the mean values $\overline{{\cal T}}^{\text{(vac)}}_{\text{(par)}}\left(\beta,x\right)$,
	and the dashed lines show ${\cal T}^{\text{(cas)}}\approx-0.023$.
	In (a) and (b), we have  $\tau=3\times10^{-1}$ and $\tau=10^{-1}$, respectively. 
	In (a), $\overline{{\cal T}}_{\text{(par)}}^{\text{(vac)}}\left(\beta,x\right)\approx-0.016$,
	whereas in (b) $\overline{{\cal T}}_{\text{(par)}}^{\text{(vac)}}\left(\beta,x\right)\approx 0.033$.
	}
	\label{fig:T}
\end{figure}

Now, we investigate the missing Casimir energy $\Delta E_{\text{(vac)}}^{\text{(vac)}}(t,\tau)$ for $\tau\neq 0$.
In these cases (and for any $\tau>0$), the length of the vacuum regions can vary in time, and, consequently, the vacuum energies  $E_{\text{(vac)}}^{\text{(vac)}}(t,\tau)$ can also vary in these regions. 
This can be calculated directly using the formulas for ${\cal T}_{A}$,...,${\cal T}_{D}$, and the equations for the null lines in Fig. \ref{fig:FD-suave} (see Appendix \ref{ap:vaccum-energy}). 
Let us consider $L_0=1$, $\alpha=0.55$, and $\beta=2.56$.
In Fig. \ref{fig:Delta-E}, the dash-dotted and dashed lines  correspond
to the cases $\tau=10^{-1}$ and $\tau=3\times10^{-1}$, respectively. 
One can see that
$\Delta E_{\text{(vac)}}^{\text{(vac)}}(t,\tau)<0$,
so that an amount of positive vacuum energy
$-\Delta E_{\text{(vac)}}^{\text{(vac)}}(t,\tau)$
needs to be found in the particle regions
[darker regions in Fig. \ref{fig:T}], added to $E^{\text{(cas)}}_{\text{(par)}}(t,\tau)$.
This, in turn, implies a mean energy density $\overline{{\cal T}}_{\text{(par)}}^{\text{(vac)}}\left(t,x\right)>{\cal T}^{\text{(cas)}}$. 

In Fig. \ref{fig:T}, we show the behavior of  
${\cal T}_{\text{(vac)}}^{\text{(vac)}}\left(t,x\right)$,
$\overline{{\cal T}}^{\text{(vac)}}_{\text{(par)}}\left(t,x\right)$,
and ${\cal T}^{\text{(cas)}}$, for $t=\beta=2.56$. 
In Figs. \ref{fig:T-3-10-1} and \ref{fig:T-10-1}, we have  $\tau=3\times10^{-1}$ and $\tau=10^{-1}$, respectively. 
Thus, in the particle regions, there is the presence of a vacuum energy greater
than the corresponding out-Casimir ones. 
However, note that $\overline{{\cal T}}_{\text{(par)}}^{\text{(vac)}}\left(\beta,x\right)$
remains negative in Fig. \ref{fig:T-3-10-1}, whereas it becomes positive in Fig. \ref{fig:T-10-1}.
In the limit $\tau\to 0$, 
$\overline{{\cal T}}_{\text{(par)}}^{\text{(vac)}}\to\infty$,
so that $\overline{{\cal T}}_{\text{(par)}}^{\text{(vac)}}$ can
be written in terms of Dirac delta functions, as already discussed.

\section{What is behind the positive vacuum energy} 
\label{sec:origin}

Let us start splitting $E_{\cal \tilde{A}}$ and $E_{\cal \tilde{B}}$
as follows: $E_{\cal \tilde{A}}
=E_{\cal \tilde{A}\text{(vac)}}^{\text{(vac)}}
+E_{\cal \tilde{A}\text{(par)}}^{\text{(vac)}}
+E_{\cal \tilde{A}}^{\text{(par)}}$;
$E_{\cal \tilde{B}}
=E_{\cal \tilde{B}\text{(vac)}}^{\text{(vac)}}
+E_{\cal \tilde{B}\text{(par)}}^{\text{(vac)}}
+E_{\cal \tilde{B}}^{\text{(par)}}$.
The energy $E_{{\cal \tilde{A}}}$ is negative
[from Eqs. \eqref{eq:T-A-tilde}, \eqref{eq:A-tilde}, and \eqref{eq:f-A}],
and it involves only
the functions ${\cal A}$ [Eq. \eqref{eq:f-A}], which, by means of a Doppler-shifted reflection of the incoming vacuum in-fields,
only contributes to the rearrangement of the vacuum energy,
giving no contribution to the particle production
($E_{\cal \tilde{A}}^{\text{(par)}}=0$).
The energy $E_{\cal \tilde{B}}$ involves the
functions ${\cal B}$ [Eq. \eqref{eq:f-B}], which are zero in vacuum regions
[$\ddot{L}(t_i)=0$ and $\dddot{L}(t_i)=0$ in this case], so that $E_{\cal \tilde{B}\text{(vac)}}^{\text{(vac)}}=0$.
Then, we have (see Sec. \ref{sec:fundamental-formulas})
$E_{\cal \tilde{A}\text{(vac)}}^{\text{(vac)}}
+E_{\cal \tilde{A}\text{(par)}}^{\text{(vac)}}+
E_{\cal \tilde{B}\text{(par)}}^{\text{(vac)}}=E^{\text{(cas)}}$,
and
$E_{\cal \tilde{B}}^{\text{(par)}}
=E^{\text{(par)}}$.

From Eqs. \eqref{eq:A-tilde} and \eqref{eq:f-A},
one can see that $E_{\cal \tilde{A}\text{(par)}}^{\text{(vac)}}<0$, and,
taking $L_0\to\infty$ (free space), 
we have: $E_{\cal \tilde{A}\text{(vac)}}^{\text{(vac)}}\to 0$,
$E_{\cal \tilde{A}\text{(par)}}^{\text{(vac)}}\to 0$, and
$E^{\text{(cas)}}\to 0$ (the vacuum energy in free space is
renormalized to zero); we also have $E_{\cal \tilde{B}\text{(par)}}^{\text{(vac)}}\to 0$.
From Eqs. \eqref{eq:B-tilde} and \eqref{eq:f-B},
one can see that $E_{\cal \tilde{B}}$ does not depend on $L_0$, so that it does not matter whether we consider a cavity
or a free space, it has the same value.
Thus, for $L_0\to\infty$, we have
$E_{\cal \tilde{B}}=E_{\cal \tilde{B}}^{\text{(par)}}\vert_{L_0\to\infty}=\int_{0}^{\infty}d\omega\omega{\cal N}\left(\omega\right),
$
where ${\cal N}\left(\omega\right)$
is the mean density of number of particles in the mode $\omega$.
%

In the particle regions in Fig. \ref{fig:T-10-1}, we have
a positive vacuum energy, this means that
$E_{\cal \tilde{A}\text{(par)}}^{\text{(vac)}}+ E_{\cal \tilde{B}\text{(par)}}^{\text{(vac)}}>0$.
Since $E_{\cal \tilde{A}\text{(par)}}^{\text{(vac)}}<0$,
the only way to achieve this is with $E_{\cal \tilde{B}\text{(par)}}^{\text{(vac)}}>0$.
In other words, it is the presence of
$E_{\cal \tilde{B}\text{(par)}}^{\text{(vac)}}>0$ that 
is behind the positive vacuum energy in the particle regions shown in Fig. \ref{fig:T-10-1}.

Putting the ideas of the last two paragraphs together, we have
that when there is only a single mirror moving in a free space,
all the energy $E_{\cal \tilde{B}}$ transferred (in this case, dissipated)
from the mirror to the field is converted into real particles [left hand side of Eq. \eqref{eq:B-E-N-L0}]. On the other hand, 
when considering a second and static mirror forming a cavity, and the same movement of the first mirror dissipating the same amount of
energy $E_{\cal \tilde{B}}$, we have that only a part of this energy is converted into real particles [right hand side of Eq. \eqref{eq:B-E-N-L0}], with the difference [$E_{\cal \tilde{B}\text{(par)}}^{\text{(vac)}}$] being converted into positive vacuum energy propagating together with these particles [right hand side of Eq. \eqref{eq:B-E-N-L0}]. Thus, we have
\begin{equation}
	\int_{0}^{\infty}d\omega\omega{\cal N}\left(\omega\right)=
	\sum_{n}\omega_n{N}_n+E_{\cal \tilde{B}\text{(par)}}^{\text{(vac)}},
	\label{eq:B-E-N-L0}
\end{equation}
with the left and right hand sides of this equation being illustrated in Figs. \ref{fig:visual-livre}
and \ref{fig:visual-cavidade}, respectively.

\section{Application to a concrete law of motion} 
\label{sec:app}

Although we
have worked with exact values for the vacuum energy density 
in vacuum regions (lighter regions in Fig. \ref{fig:T}),
we only worked with mean values of the vacuum energy density in the particle regions (darker regions in Fig. \ref{fig:T}),
and nothing has been said about the particle energy itself. 
In this section, we discuss the exact behavior of ${\cal{T}}\left(t,x\right)$, 
examining its parts ${\cal{T}}_{{\cal \tilde{A}}}\left(t,x\right)$ and ${\cal{T}}_{{\cal \tilde{B}}}\left(t,x\right)$, in whole space, including
in the particle regions.
Although ${\cal{T}}\left(t,x\right)$ is given exactly by Eq. \eqref{eq:T}, 
this formula can require a numerical treatment.
In this section, our results are found taking as basis the computer routines described in Ref. \cite{Alves-Granhen-CPC-2014}.

\begin{figure}
	\centering
	\subfigure[\label{fig:T-A-tilde-tau-10-1}]{\epsfig{file=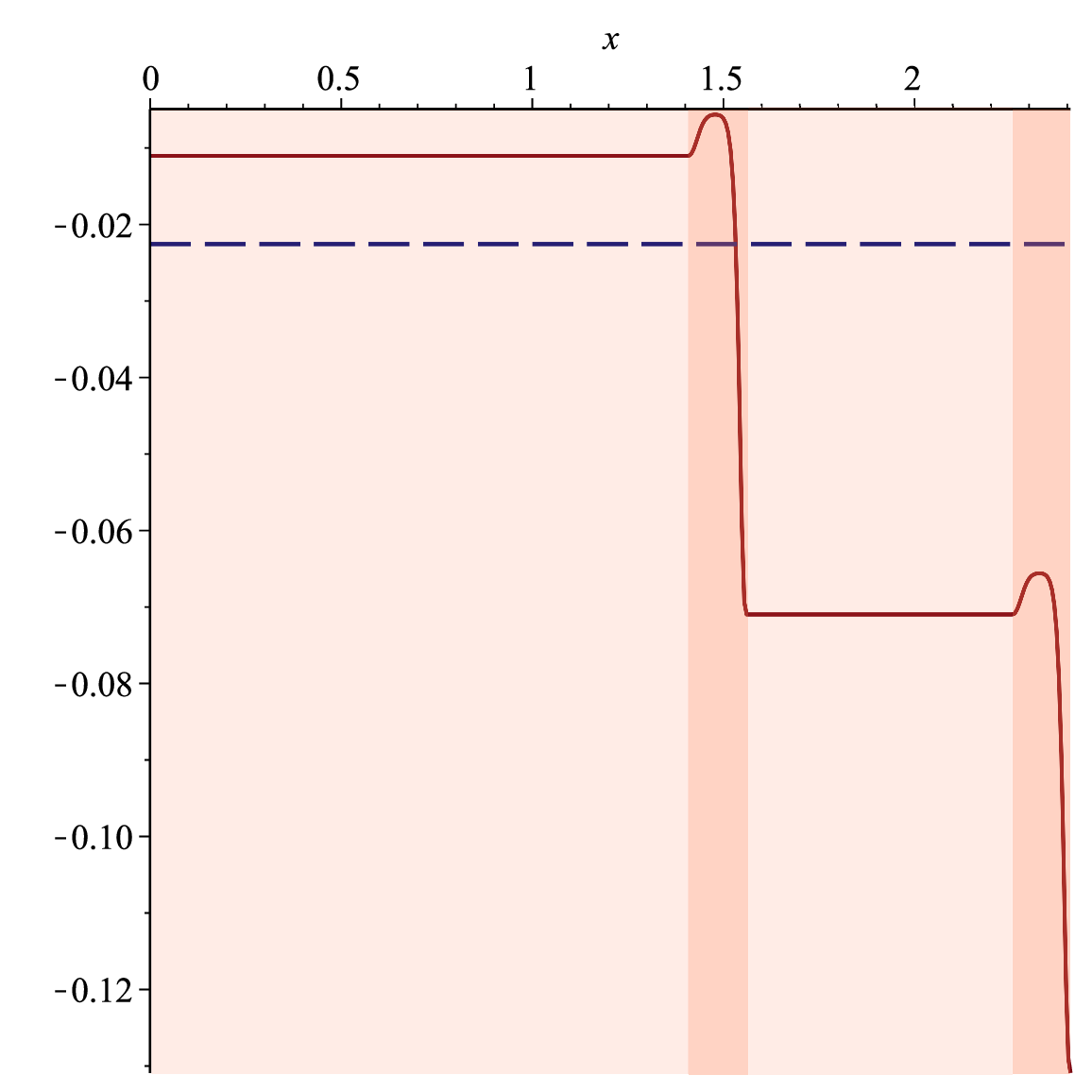, width=0.45\linewidth}}
	\subfigure[\label{fig:T-B-tilde-tau-10-1}]{\epsfig{file=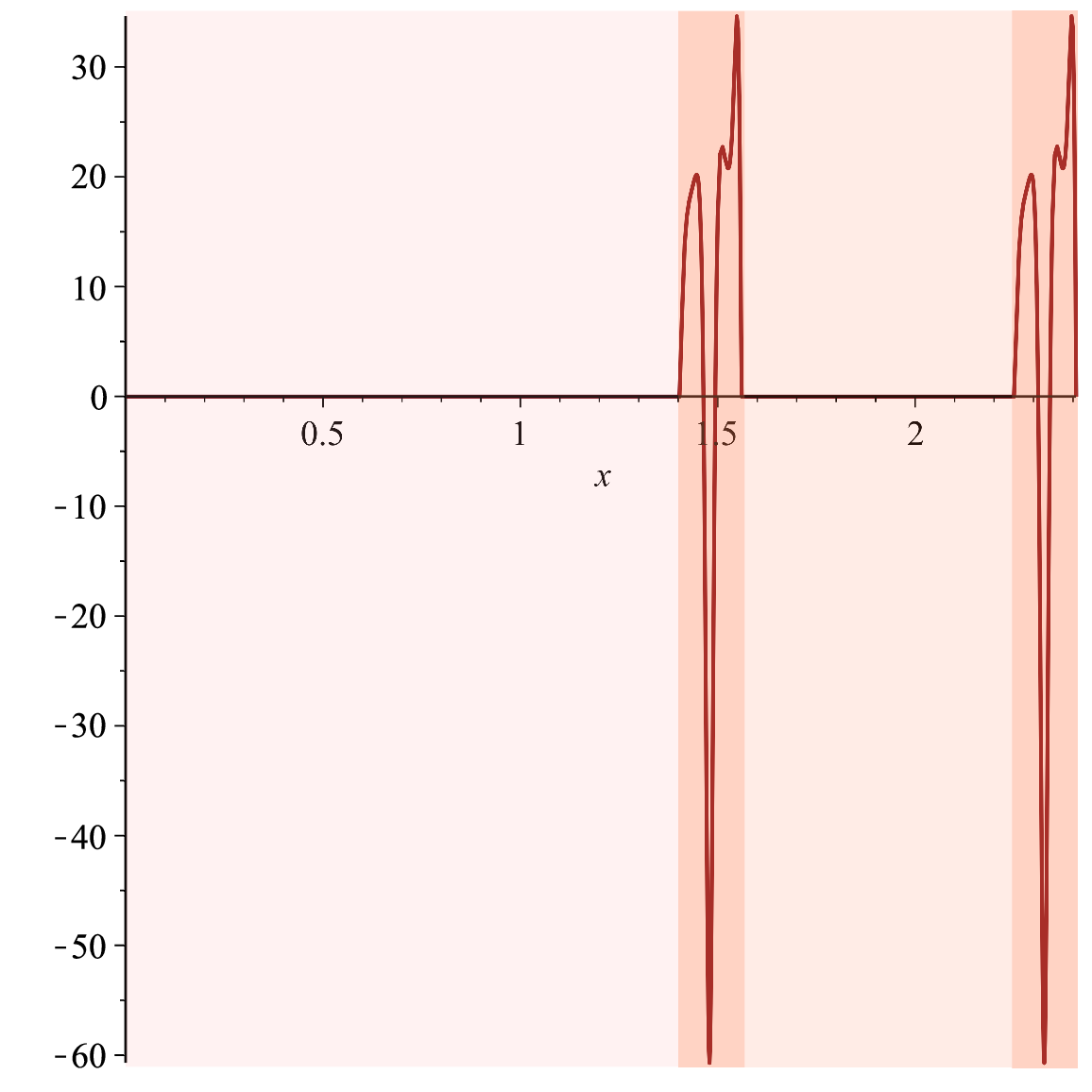, width=0.45 \linewidth}}
	\caption{
		Energy densities (vertical axes) ${\cal T}_{{\cal \tilde{A}}}$ [shown in (a)] and 
		${\cal T}_{{\cal \tilde{B}}}$ [in (b)], at $t=\beta$, versus $x$ (horizontal axes), for
		$0\leq x\leq L_0+\alpha\beta$, considering $L_0=1$, $\alpha=0.55$, $\beta=2.56$, and $\tau=10^{-1}$.
		In both figures, lighter regions are vacuum regions, whereas darker regions are particle regions.
	}
	\label{fig:T-A-tilde-B-tilde-10-1}
\end{figure}

Let us focus on the case corresponding to
Fig. \ref{fig:T-10-1}, where $L_0=1$, $\alpha=0.55$, $\beta=2.56$, and
$\tau=10^{-1}$. 
Our calculations show that, for these values, the functions $L_1(t)$ and $L_2(t)$ have the following parameters (see Appendix \ref{ap:smoothing}): 
$a_{1}\approx 10999.96$; $b_{1}\approx-2.48\times 10^5$; 
$c_{1}\approx 1.98\times 10^{6}$; $d_{1}\approx-5.50\times 10^{6}$; $a_{2}\approx -5.50\times 10^{6}$; 
$b_{2}\approx -1.98\times 10^{6}$; $c_{2}\approx -2.48 \times 10^5$; $d_{2}\approx -11000.04$.
We show the exact behavior of ${\cal{T}}_{{\cal \tilde{A}}}\left(\beta,x\right)$ in Fig. \ref{fig:T-A-tilde-tau-10-1}, and ${\cal{T}}_{{\cal \tilde{B}}}\left(\beta,x\right)$ in Fig. \ref{fig:T-B-tilde-tau-10-1}.

After integrating the energy densities in Fig. \ref{fig:T-A-tilde-B-tilde-10-1},
in each of their respective regions, and using previous formulas, we obtain that (see Appendix \ref{ap:details-application}): 
$E_{\cal \tilde{B}}\left(\beta\right)\approx 2.699$,
$E_{\cal \tilde{B}\text{(par)}}^{\text{(vac)}}\left(\beta\right)\approx 0.025$,
and $E_{\cal\tilde{B}}^{\text{(part)}}\left(\beta\right)\approx 2.674$.
Thus, as discussed in Sec. \ref{sec:origin},
here we find that not all of the energy $E_{\cal \tilde{B}}$ is converted into 
the energy $E_{\cal\tilde{B}}^{\text{(part)}}$ of real particles, with the difference being converted into 
the positive vacuum energy $E_{\cal \tilde{B}\text{(par)}}^{\text{(vac)}}$, which is located in the same regions where the particles can be.
From our calculations we also recover (see Appendix \ref{ap:details-application}) $\overline{{\cal T}}_{\text{(par)}}^{\text{(vac)}}\left(t,x\right)
\approx0.033$, as indicated in Fig. \ref{fig:T-10-1}.
The ratio $E_{\cal \tilde{B}\text{(par)}}^{\text{(vac)}}(\beta)/E_{\cal \tilde{B}}\approx 0.9\%$,
so $0.9\%$ of the energy $E_{\cal \tilde{B}}$ dissipated from the mirror is converted into vacuum
positive energy.
The ratio $[E_{\cal \tilde{A}\text{(par)}}^{\text{(vac)}}+E_{\cal \tilde{B}\text{(par)}}^{\text{(vac)}}]/[
E_{\cal \tilde{A}\text{(par)}}^{\text{(vac)}}+E_{\cal \tilde{B}}]\approx 0.4\%$, which means that an average value of
$0.4\%$ of the total energy in the particle regions is positive vacuum energy.

\section{Final remarks} 

We showed that, in addition to real particles, the excitation of the quantum vacuum in a dynamical cavity can also produce a cloud of positive vacuum energy around these particles. 
In the same way that the creation of particles by the excitation of the quantum vacuum,
as predicted by Moore
\cite{Moore-1970}, has been also predicted in the context of other fields, boundary conditions, and spacetime dimensions \cite{Dodonov-Phys-2020}, we expect that the possibility of creation of positive vacuum energy around particles, predicted here, can also be generalized to any confined field under dynamic boundary conditions.  
Moreover, since, in addition to moving mirrors,
models with non-static spacetime (expanding universe) are another way
to disturb the quantum vacuum and create particles \cite{Davies-JOPTB-2005},
it is also interesting to investigate if such a vacuum positive energy can
be present around massive particles created in these models, and 
if the equivalent in mass of this vacuum energy could be of any relevance to the understanding of some aspect of the dark matter around the visible matter.
Finally, considering that some experiments have
already observed photons created via excitation of the quantum vacuum
\cite{Wilson-Nature-2011,Lahteenmaki-2013, Vezzoli-2019,Schneider-et-al-PRL-2020},
an observation of a signature of the 
vacuum positive energy predicted here, 
namely the difference
between the energy of the spectrum of created particles in
a free space and in a cavity [Eq. \eqref{eq:B-E-N-L0}], seems not unfeasible.

\begin{acknowledgments}
The author thanks: A. L. C. Rego, A. N. Braga, C. Farina, 
E. R. Granhen, J. D. L. Silva, L. Oliveira,
L. Queiroz, O. J. N. Neto, P. A. M. Neto, S. S. Coelho, T. R. Alves,
T. R. Alves, V. S. Alves, and V. V. Dodonov for valuable discussions; 
A. L. C. Rego, E. R. Granhen, J. D. L. Silva, L. Queiroz, and V. S. Alves for 
a careful reading of this manuscript; 
L. Queiroz and O. J. N. Neto for the support in improving some figures presented here.
V. V. Dodonov and A. V. Dodonov for the invitation which enabled the presentation of 
a preliminary version of this work in the Second International Workshop on Nonstationary Systems - Brasília (2023),
with support of the FAPDF - Brazil;
E. C. Marino for the invitation which enabled the presentation of 
another version of this work in the workshop Quantum Field Theory Approach to Condensed Matter Physics - Rio de Janeiro (2024).
%
This work was partially supported by CNPq - Brazil, Processo 408735/2023-6 CNPq/MCTI.
\end{acknowledgments}

\appendix
\section{Dynamical and static cavities: basic formulas}
\label{ap:basic-formulas}

The real scalar field field operator, solution of the wave equation
in 1+1 dimensions, is given by (see, for instance, Ref. \cite{Moore-1970,Fulling-Davies-PRSA-1976,Alves-Granhen-Silva-Lima-PRD-2010})
\begin{equation}
	\hat{\phi}(t,x)=\sum^{\infty }_{n=1}\left[
	\hat{a}^{\text{(in)}}_{n}\phi_{n}\left( t,x\right) +H.c.\right],
	\label{eq:field-solution-1}
\end{equation} 
where $\hat{a}_{n}^{{\text{(in)}}\dag}$ and $\hat{a}^{\text{(in)}}_{n}$ are, respectively, the creation and annihilation
operators related to the $n$th field mode in the cavity with length $L_0$,
and the field modes $\phi_{n}(t,x)$ are given by
\begin{eqnarray}
	\phi_{n}(t,x)=\frac{i}{2\sqrt{n\pi}}\left[\varphi_{n}(v)
	-\varphi_{n}(u)\right],
	\label{eq:field-solution-2}
\end{eqnarray} 
with $\varphi_{n}(z)=e^{-in\pi R(z)}$, $u=t-x$, $v=t+x$, and $R$ satisfying Moore's functional equation
\begin{equation}
	R[t+L(t)]-R[t-L(t)]=2.
	\label{eq:Moore-equation}
\end{equation}
The exact formulas for ${\cal T}(t,x)$, the expected value of the energy 
density operator $\hat{T}_{00}(t,x)$, so that ${\cal T}(t,x)=\langle\hat{T}_{00}(t,x)\rangle_{\text{in}}$,
can be split as (see, for instance, Refs. \cite{Alves-Granhen-Silva-Lima-PRD-2010,Alves-Granhen-CPC-2014})
%
%
\begin{equation}
	{\cal T}(t,x)={\cal T}_{\text{from-vac}}(t,x)+ {\cal T}_{\text{from-par}}(t,x),
	\label{eq:T-Apendice}
\end{equation}
where 
\begin{gather}
	{\cal T}_{\text{from-vac}}(t,x)=\frac{\pi}{4}\sum_{n=1}^{\infty}n\left[R^{\prime2}\left(v\right)+R^{\prime2}\left(u\right)\right], \label{eq:vac}
\end{gather}
and 
\begin{equation}
	{\cal T}_{\text{ from-par}}(t,x)={\cal T}_{\left\langle \hat{a}^{{\text{(in)}}\dag}\hat{a}^{\text{(in)}}\right\rangle_{\text{in}}}(t,x)+
	{\cal T}_{\left\langle \hat{a}^{\text{(in)}}\hat{a}^{\text{(in)}}\right\rangle_{\text{in}}}(t,x),
\end{equation} 
with
\begin{gather}
	{\cal T}_{\left\langle \hat{a}^{{\text{(in)}}\dag}\hat{a}^{\text{(in)}}\right\rangle_{\text{in}}}(t,x)
	=g_{1}(v) + g_{1}(u),\label{eq:T-a-adaga-a}\\
	{\cal T}_{\left\langle \hat{a}^{\text{(in)}}\hat{a}^{\text{(in)}}\right\rangle_{\text{in}}}(t,x)=g_{2}(v) + g_{2}(u),
	\label{eq:T-a-a}
\end{gather}
%
%
\begin{gather}
	g_{1}(z) = \frac{\pi}{2}\sum_{n,n^{\prime}=1}^{\infty }
	\sqrt{ n n^{\prime }}  
	\nonumber \\
	\times{\mbox Re} \left\{e^{i\left(
	n-n^{\prime }\right) \pi R\left(z\right) }\left[R^{\prime }\left(z\right)
	\right]^{2} \left\langle \hat{a}^{\text{(in)}\dag}_{n}\hat{a}^{\text{(in)}}_{n^{\prime }}\right\rangle_{\text{in}}\right\},
	\label{eq:g1}
\end{gather}
\begin{gather}
	g_{2}(z) =\frac{\pi}{2}\sum_{n,n^{\prime }=1}^{\infty }\sqrt{ n  n^{\prime }
	}  \nonumber\\
	\times{\mbox Re} \left\{e^{-i\left( n+n^{\prime }
		\right) \pi R\left( z\right) }\left[ R^{\prime }\left( z\right) \right]
	^{2} \left\langle \hat{a}^{\text{(in)}}_{n}\hat{a}^{\text{(in)}}_{n^{\prime }}\right\rangle_{\text{in}}
	\right\}.   
	\label{eq:g2}
\end{gather}
%
The averages $\langle...\rangle_{\text{in}}$,
are here taken over any initial field state (further on, 
as mentioned in Sec. \ref{sec:fundamental-formulas}, we apply our calculation considering the vacuum as the initial field state, the focus of the present paper).
The term ${\cal T}_{\text{from-vac}}$ has its origin in the vacuum fluctuations, whereas ${\cal T}_{\text{ from-par}}$ in the presence of real particles in the initial field state [see Eqs. \eqref{eq:vac}-\eqref{eq:g2}].

The term ${\cal T}_{\text{ from-vac}}$ is divergent, but 
adopting the point-splitting regularization method \cite{Fulling-Davies-PRSA-1976} one has
${\cal T}_{\text{ from-vac}}$, now redefined as the renormalized local energy density, 
given by \cite{Fulling-Davies-PRSA-1976} 
%
\begin{equation}
	{\cal T}_{\text{ from-vac}}(t,x) = -f(v) -f(u), 
	\label{eq:T-vac-ren}
\end{equation}
where 
\begin{equation}
	f=\frac{1}{24\pi}\left[\frac{R^{\prime\prime\prime}}{R^{\prime}}-\frac{3}
	{2}\left(\frac{R^{\prime\prime}}{R^{\prime}}\right)^{2}
	+\frac{\pi^{2}}{2}{R^{\prime}}^{2}\right],
	\label{eq:f}
\end{equation}
with the derivatives taken with respect to the argument of the $R$ function.
The value of the energy density at each point of the spacetime [Eq. \eqref{eq:T}], can be calculated recursively, as discussed in Ref. \cite{Alves-Granhen-Silva-Lima-PRD-2010} and shown in Eqs. \eqref{eq:T}-\eqref{eq:f-B}.

For a static ``in'' situation with an arbitrary initial field state, 
we can write the total energy density ${\cal T}(t,x)$, 
now relabeled as ${\cal T}^{\text{(in)}}(t,x)$,
as given by (see Ref. \cite{Alves-Granhen-Silva-Lima-PRD-2010})
\begin{equation}
	{\cal T}^{\text{(in)}}(t,x)=
	{\cal T}^{\text{(cas-in)}}+{\cal T}^{\text{(in)}}_{\left\langle \hat{a}^{{\text{(in)}}\dag}\hat{a}^{\text{(in)}}\right\rangle_{\text{in}}}(t,x)+{\cal T}^{\text{(in)}}_{\left\langle \hat{a}^{\text{(in)}}\hat{a}^{\text{(in)}}\right\rangle_{\text{in}}}(t,x),
	\label{eq:T-s}
\end{equation}
where ${\cal T}_{\text{cas}}^{\text{(in)}}=-{\pi}/({24L_0^2})$ is the Casimir energy density, and the functions ${\cal T}^{\text{(in)}}_{\left\langle \hat{a}^{{\text{(in)}}\dag}\hat{a}^{\text{(in)}}\right\rangle_{\text{in}}}(t,x)$ and ${\cal T}^{\text{(in)}}_{\left\langle \hat{a}^{\text{(in)}}\hat{a}^{\text{(in)}}\right\rangle_{\text{in}}}(t,x)$
are obtained by using $R(z)={z}/{L_0}$ in Eqs. \eqref{eq:T-a-adaga-a}
and \eqref{eq:T-a-a}, respectively.
%
%
For an arbitrary field state, we have \cite{Alves-Granhen-Silva-Lima-PRD-2010}:
\begin{gather}
	\int_0^{L_0} {\cal T}^{\text{(cas-in)}}\;dx=E^{\text{(cas-in)}},\;\;
	\int_0^{L_0}{\cal T}_{\left\langle \hat{a}^{\text{(in)}}\hat{a}^{\text{(in)}}\right\rangle_{\text{in}}}(t,x)\;dx=0,\label{eq:int-E-cas}\\
	\int_0^{L_0}{\cal T}_{\left\langle \hat{a}^{{\text{(in)}}\dag}\hat{a}^ {\text{(in)}}\right\rangle_{\text{in}}}(t,x)\;dx=\sum_{n=1}^{\infty }\omega_{n}^{\text{(in)}}{N}^{\text{(in)}}_n,
	\label{eq:int-g2}	
\end{gather}
where the Casimir energy is $E^{\text{(cas-in)}}=-{\pi}/({24L_0})$, $\omega_{n}^{\text{(in)}}=n\pi/L_0$, and
${N}_n^{\text{(in)}}=\left\langle \hat{a}_{n}^{\text{(in)}\dag }\hat{a}^{\text{(in)}}_{n}\right\rangle_{\text{in}}$ is the 
mean number of particles in the $n$th mode.
We highlight that the function ${\cal T}_{\left\langle \hat{a}^{\text{(in)}}\hat{a}^{\text{(in)}}\right\rangle_{\text{in}}}(t,x)$
can give contribution for the local behavior of the energy density ${\cal T}(t,x)$, but not for the total energy stored in the cavity. 
From Eqs. \eqref{eq:T-s}-\eqref{eq:int-g2}, the total energy $E^{\text{(in)}}=
\int_0^{L_0}{\cal{T}}(t,x)\;dx$ can be written as \cite{Alves-Granhen-Silva-Lima-PRD-2010}
\begin{equation}
	E^{{\text{(in)}}}=E^{\text{(cas-in)}}+E^{\text{(par-in)}},
	\label{eq:E-s}
\end{equation}
where $E^{\text{(par-in)}}=\sum_{n}\omega_{n}^{{\text{(in)}}}{N}_n^{\text{(in)}}$ is the total energy corresponding to the real particles.
We highlight that, independently of the field state, the total vacuum energy in the static cavity is the Casimir one. 
Also note that, despite the fact that ${\cal T}_{\left\langle \hat{a}^{{\text{(in)}}\dag}\hat{a}^{\text{(in)}}\right\rangle_{\text{in}}}$ and ${\cal T}_{\left\langle \hat{a}^{\text{(in)}}\hat{a}^{\text{(in)}}\right\rangle_{\text{in}}}$ can depend on time, from the principle
of energy conservation, the total energy is a constant in time for the static situation.

When the mirror stops its motion at a instant $\beta$, with the cavity reaching a final length $L_{\text{fin}}$, we have an ``out'' state, in general different from vacuum (even when considering the vacuum as the initial field state).
In the ``out'' state, the field operator $\hat{\phi}(t,x)$ can be written
as in Eq. \eqref{eq:field-solution-1} or as (see,
for instance, Refs. \cite{Dodonov-Klimov-Manko-PLA-1990,Dodonov-JMP-1993})
\begin{equation}
	\hat{\phi}(t,x)=\sum^{\infty }_{n=1}\left[
	\hat{a}_{n}\phi^{(0)}_{n}\left( t,x\right) +H.c.\right],
	\label{eq:field-solution-out}
\end{equation} 
where $\hat{a}_{n}^{\dag}$ and $\hat{a}_{n}$ are, respectively, the creation and annihilation
operators related to the $n$th field mode in the cavity with length $L_{\text{fin}}$,
and the field modes $\phi^{(0)}_{n}\left( t,x\right)$ are obtained
by replacing $R(z)\to z/L_{\text{fin}}$ in Eq. \eqref{eq:field-solution-2}.
The energy density ${\cal T}(t,x)$, 
now relabeled as ${\cal T}^{\text{(out)}}(t,x)$,
is given by (see Ref. \cite{Alves-Granhen-Silva-Lima-PRD-2010})
\begin{equation}
	{\cal T}^{\text{(out)}}(t,x)=
	{\cal T}^{\text{(cas)}}+{\cal T}^{\text{(out)}}_{\left\langle \hat{a}^{\dag}\hat{a}\right\rangle_{\text{in}}}(t,x)+{\cal T}^{\text{(out)}}_{\left\langle \hat{a}\hat{a}\right\rangle_{\text{in}}}(t,x),
	\label{eq:T-s-out}
\end{equation}
where ${\cal T}^{\text{(cas)}}=-\pi/(24L_{\text{fin}}^2)$ is the Casimir energy density, and the functions ${\cal T}^{\text{(out)}}_{\left\langle \hat{a}^{\dag}\hat{a}\right\rangle_{\text{in}}}(t,x)$ and ${\cal T}^{\text{(out)}}_{\left\langle \hat{a}\hat{a}\right\rangle_{\text{in}}}(t,x)$
are obtained from Eqs. \eqref{eq:T-a-adaga-a} and \eqref{eq:T-a-a}, respectively,
by making in Eqs. \eqref{eq:g1} and \eqref{eq:g2}: $R(z)={z}/{L_{\text{fin}}}$, 
$\hat{a}_{n}^{{\text{(in)}}\dag}\to\hat{a}_{n}^{\dag}$, 
and $\hat{a}_{n}^{\text{(in)}}\to\hat{a}_{n}$.
%
After performing integrations similar to
those in Eqs. \eqref{eq:int-E-cas} and \eqref{eq:int-g2},
we find that the total energy in the cavity in ``out'' situation is given as 
written in  Sec. \ref{sec:fundamental-formulas}.

\section{Recovering Fulling-Davies formulas}
\label{ap:recovering-FD}

The particular case of a dynamical cavity subjected to the law of motion
given in Eq. \eqref{eq:FD-suave} with $\tau=0$, with $\beta=2L_{0}/\left(1-\alpha\right)$, was
considered by Moore \cite{Moore-1970}. In this case, the
null line corresponding to energy density produced by the discontinuity in the mirror velocity at $t=0$,
after reflection from the static mirror at $x=0$, intersects the moving mirror
exactly when it instantaneously decelerates, which implies in no particle
creation in this process \cite{Moore-1970,Fulling-Davies-PRSA-1976,Castagnino-Ferraro-AnnPhys-1984}.
Here, our focus is on situations where $\beta<2L_{0}/\left(1-\alpha\right)$, which means that particles can be created
\cite{Fulling-Davies-PRSA-1976,Castagnino-Ferraro-AnnPhys-1984}.

Due to the presence, in the law of motion given by Eq. \eqref{eq:FD-suave} with $\tau=0$, of sudden jumps in the derivatives of $L(t)$ at $t=0$ and $t=\beta$, from Eq. \eqref{eq:T}, one can see that the energy density is not defined on the null lines [dashed lines in Fig.\ref{fig:FD}].
%
By means of the same equations, we recover the vacuum energy densities in regions $A$, $B$, $C$ and $D$, as found in Ref. \cite{Fulling-Davies-PRSA-1976}.
At any point $(t,x)$ in the region $A$, the function $f(u)=\pi/(48L_0^2)$, and 
$f\left(v\right)=\left[\pi/(48L_0^2)\right]\left[\left(1-\alpha\right)/\left(1+\alpha\right)\right]^{2}$.
Note that this expression for $f\left(v\right)$ 
can be viewed as the Doppler-shifted reflection of the incoming
constant field $f(u)=f^{(s)}=\pi/(48L_0^2)$, which is modified by the factor 
$\left[\left(1-\alpha\right)/\left(1+\alpha\right)\right]^{2}$ \cite{Cole-Schieve-PRA-2001}.
Also note that ${\cal B}(t)=0$, since there is no acceleration in the mirror.
Thus, using these expressions for $f(u)$ and $f(v)$ in Eq. \eqref{eq:T-vac-ren}, we recover the vacuum energy density ${\cal T}_{A}$ given in Sec. \ref{sec:missing-and-positive}.
In region $B$, the function $f(v)=\pi/(48L_0^2)$, and 
$f\left(u\right)=\left[\pi/(48L_0^2)\right]\left[\left(1-\alpha\right)/\left(1+\alpha\right)\right]^{2}$.
Thus, we recover ${\cal T}_{B}={\cal T}_{A}$.
In region $C$, we have $f(u)=f(v)=\pi/(48L_0^2)$, so that ${\cal T}_{C}={\cal T}^{\text{(cas,in)}}=-\pi/(24L_0)$,
as shown in Sec. \ref{sec:missing-and-positive}.
In region $D$, we have $f(u)=f(v)=\left[\pi/(48L_0^2)\right]\left[\left(1-\alpha\right)/\left(1+\alpha\right)\right]^{2}$, 
and thus, using Eq. \eqref{eq:T-vac-ren}, we recover ${\cal T}_{D}$ given in Sec. \ref{sec:missing-and-positive}.

\section{Expanding cavities with continuous derivatives of ${L}(t)$}
\label{ap:smoothing}

One way to avoid the divergences in the energy density occurred
at $t=0$ and $t=\beta$ when considering the law of motion 
as shown in Fig. \ref{fig:FD}, is looking for a law of motion such that:
\begin{equation}
	L\left(t\right)=\begin{cases}
		L_0 & \left(t<0\right),\\
		L_{1}\left(t\right) & \left(0\leq t \leq \tau_{1}\right),\\
		L_{1}\left(\tau_{1}\right)+\alpha\left(t-\tau_{1}\right) & \left(\tau_{1}<t<\beta-\tau_{2}\right),\\
		L_{2}\left(t\right) & \left(\beta-\tau_{2}\leq t \leq \beta\right),\\
		L_{\text{fin}}& \left(\beta<t\right),
	\end{cases}
	\label{eq:FD-suave-super-geral}
\end{equation}
where $\dot{L}_{1}\left(t\right)$, $\ddot{L}_{1}\left(t\right)$,
$\dddot{L}_{1}\left(t\right)$, $\dot{L}_{2}\left(t\right)$, $\ddot{L}_{2}\left(t\right)$,
and $\dddot{L}_{2}\left(t\right)$ are well defined functions, and:
\begin{align}
	&L_{1}\left(0\right)=L_0;\ \dot{L}_{1}\left(0\right)=\ddot{L}_{1}\left(0\right)
	=\dddot{L}_{1}\left(0\right)=0;
	\label{eq:conditions-geral-L1-0}\\
	&\dot{L}_{1}\left(\tau_{1}\right)=\alpha;\;\ddot{L}_{1}\left(\tau_{1}\right)=\dddot{L}_{1}\left(\tau_{1}\right)=0;
	\label{eq:conditions-geral-L1-t1}\\
	&L_{2}\left(\beta-\tau_{2}\right)=L_{1}\left(\tau_{1}\right)+\alpha\left(\beta-\tau_{2}-\tau_1\right);\ \label{eq:conditions-geral-L2-t2-1}\\
	&\dot{L}_{2}\left(\beta-\tau_{2}\right)=\alpha;\ddot{L}_{2}\left(\beta-\tau_{2}\right)=\dddot{L}_{2}\left(\beta-\tau_{2}\right)=0;\;\label{eq:conditions-geral-L2-t2-2}\\
	&L_{2}\left(\beta\right)=L_{\text{fin}};\ \dot{L}_{2}\left(\beta\right)=\ddot{L}_{2}\left(\beta\right)=\dddot{L}_{2}\left(\beta\right)=0.
	\label{eq:conditions-geral-L2-beta}
\end{align}
These conditions guarantee that the energy density given in Eq. \eqref{eq:T}
is well defined in all spacetime points.

Now, for simplicity, we choose
\begin{align}
	&\tau_1=\tau_2=\tau<\beta/2,
	\label{eq:tau}\\
	&L_1(\tau)=L_0+\alpha\tau,
	\label{eq:conditions-L1-especifica}\\
	&L_{\text{fin}}=L_0+\alpha\beta.
	\label{eq:conditions-L2-especifica}
\end{align}
Thus, we propose $L(t)$ as given in Eq. \eqref{eq:FD-suave}.
The function $L_{1}$ naturally satisfies Eq. \eqref{eq:conditions-geral-L1-0}, and
$L_{2}$ satisfies Eqs. \eqref{eq:conditions-geral-L2-beta} and \eqref{eq:conditions-L2-especifica}.
The constants $a_1$,  $b_1$,  $c_1$ and $d_1$ are found
by solving Eqs. \eqref{eq:conditions-geral-L1-t1} and \eqref{eq:conditions-L1-especifica}, whereas $a_2$,  $b_2$,  $c_2$ and $d_2$, by solving Eqs. \eqref{eq:conditions-geral-L2-t2-1} and \eqref{eq:conditions-geral-L2-t2-2}.
The solution found is valid if 
\begin{equation}
	|\dot{L}\left(t\right)|<1.
	\label{eq:abs-L-menor-1}
\end{equation}
We show in Sec. \ref{sec:app} that valid solutions for $L(t)$ can be found for appropriate choices of 
$L_0$, $\alpha$, $\beta$ and $\tau$. 

\section{Vacuum and Casimir energy in vacuum regions}
\label{ap:vaccum-energy}

In this section, we focus our discussion on situations like the one illustrated in Fig. \ref{fig:FD-suave}.
Let us define the following instants: 
$t_a^{(1)}=L_0$,
$t_{a}^{\prime(1)}=L_0+\tau(1+\alpha)$,
$t_{\text{enc}}=L_0+(1/2)(\beta-\tau)(1+\alpha)$,
$t_N=L_0+(1/2)\beta(1+\alpha)$,
$t_{\text{des}}= L_0+(1/2)(\beta-\tau)(1+\alpha)+\tau(1+\alpha)$,
$t_{a}^{(2)}=2L_0+\alpha\beta$,
$t_{a}^{\prime(2)}=2L_0+\tau(1+\alpha)+\alpha\beta$,
$t_{b}^{\prime(1)}=L_0+(\beta-\tau)(1+\alpha)$,
$t_{b}^{(1)}=L_0+\beta(1+\alpha)$.
We consider $t_{b}^{\prime(1)}\geq t_{a}^{\prime(2)}$,
which 
implies that $\beta\geq L_{0}+2\tau\left(\alpha+1\right)$
(this also implies that $\beta\geq t_{a}^{\prime(1)}$).
In addition, we also consider a value of $\beta$ chosen in such a way
that in the time interval $[0,\beta]$ 
there is no superposition between the pulses of the excited field,
which implies that $\beta\leq{[2L_{0}-\tau(\alpha+1)]}/{(1-\alpha)}$
[for $\tau\neq0$, this means that $\beta<{2L_{0}}/{(1-\alpha)}$,
in agreement with our previous statement].
%
%
Both conditions lead to $L_0+2\tau(1+\alpha)\leq \beta \leq [2L_0-\tau(\alpha+1)]/(\alpha-1)$.
The formula for $E_{\text{(vac)}}^{(\text{vac})}\left(t\right)$,
for $\beta\leq t<t_{b}^{(1)}$, is given by
\begin{equation}
	E^{\text{(vac)}}_{(\text{vac})}\left(t\right) =\begin{cases}
		E^{\text{(vac)}}_{(\text{DAC})} & \left(\beta\leq t<t_{\text{enc}}\right),\\
		E^{\text{(vac)}}_{(\text{DC}1)}(t) & \left(t_{\text{enc}}\leq t<t_{N}\right),\\
		E^{\text{(vac)}}_{(\text{DC}2)}(t) & \left(t_{N}^{(1)}\leq t<t_{\text{des}}\right),\\
		E^{\text{(vac)}}_{(\text{DBC})} & \left(t_{\text{des}}\leq t<t_{a}^{(2)}\right),\\
		E^{\text{(vac)}}_{(\text{DB}1)}(t) & \left(t_{a}^{(2)}\leq t<t_{N}^{(2)}\right),\\
		E^{\text{(vac)}}_{(\text{DB}2)}(t) & \left(t_{N}^{(2)}\leq t<t_{a}^{\prime(2)}\right),\\
		E^{\text{(vac)}}_{(\text{DBD})} & \left(t_{a}^{\prime(2)}\leq t<t_{b}^{\prime(1)}\right),\\
		E^{\text{(vac)}}_{(\text{BD1})}(t) & \left(t_{b}^{\prime(1)}\leq t<t_{N}^{(2)}\right),\\
		E^{\text{(vac)}}_{(\text{BD2})}(t) & \left(t_{N}^{(2)}\leq t<t_{b}^{(1)}\right),
		\label{eq:E-out-vac-vac}
	\end{cases}
\end{equation}
where:
\begin{align}
	E^{\text{(vac)}}_{(\text{DAC})}=&\frac{\pi}{24L_{0}^{2}\left(1+\alpha\right)}
	\left[\alpha^{2}\left(\beta-2\tau\right)\right.
	\nonumber\\
	&\left.+\alpha\left(-\beta+2\tau+L_{0}\right)-2\tau+L_{0}\right],
	\label{eq:E-DAC}
\end{align}
%
%
\begin{align}
	E^{\text{(vac)}}_{(\text{DC}1)}(t)=&-\frac{\pi}{24L_{0}^{2}}\left[\frac{\left(1-\alpha\right)^{2}\left(t-L_{0}-\tau\left(1+\alpha\right)\right)}{\left(1+\alpha\right)^{2}}\right]\nonumber\\
	&
	-\frac{\pi}{24L_{0}^{2}}\left(\beta+t\right),
\end{align}
\begin{align}
	E^{\text{(vac)}}_{(\text{DC}2)}(t)=&\frac{\pi}{24L_{0}^{2}}\frac{\left(1-\alpha\right)^{2}\left[-t+L_{0}+\left(\beta-\tau\right)\left(1+\alpha\right)\right]}{\left(1+\alpha\right)^{2}}
	\nonumber\\
	&-\frac{\pi}{24L_{0}^{2}}\left(\alpha\beta+2{\it L_{0}}-t\right),
\end{align}
\begin{align}
	E^{\text{(vac)}}_{(\text{DBC})}=E^{\text{(vac)}}_{(\text{DAC})},
	\label{eq:E-DBC}
\end{align}
\begin{align}
	E^{\text{(vac)}}_{(\text{DB1})}(t)=&\frac{\pi}{24L_{0}^{2}\left(1+\alpha\right)}\left[-t\left(\alpha+1\right)
	+2\alpha^{2}\tau+L_{0}\alpha\right. \nonumber\\
	&\left.+2\alpha\beta-2\tau\alpha+L_{0}+2\tau\right],
\end{align}
\begin{align}
	E^{\text{(vac)}}_{(\text{DB2})}(t)=&-\frac{\pi\left(1-\alpha\right)^{2}}{24L_{0}^{2}\left(1+\alpha\right)^{2}}\left[-t+L_{0}+\left(\beta-\tau\right)\left(1+\alpha\right)\right]\nonumber\\
	&-\frac{\pi\left(\alpha^{2}+1\right)}{24L_{0}^{2}\left(1+\alpha\right)^{2}}\left[2L_{0}+\beta\left(-1+\alpha\right)\right],
\end{align}
\begin{align}
	E^{\text{(vac)}}_{(\text{DBD})}=&-\frac{\pi}{24L_{0}^{2}\left(1+\alpha\right)}
	\left[\left(\alpha-1\right)\alpha\beta-2\alpha\tau+L_{0}\alpha\right.\nonumber\\
	&\left.+4\tau\alpha+L_{0}-2\tau\right],
\end{align}
\begin{align}
	E^{\text{(vac)}}_{(\text{BD1})}(t)=&-\frac{\pi}{24L_{0}^{2}\left(1+\alpha\right)^{2}}
	\left[\left(\alpha^{2}-1\right)\beta-\alpha^{3}\tau+\alpha^{2}t
	\right.\nonumber\\
	&\left.+\alpha^{2}\tau+4L_{0}\alpha-2\alpha t+\tau\alpha+t-\tau\right],
\end{align}
\begin{align}
	E^{\text{(vac)}}_{(\text{BD2})}(t)=&-\frac{\pi}{24L_{0}^{2}\left(1+\alpha\right)}
	\left[\left(2\alpha^{2}+\alpha+1\right)\beta-2\alpha^{2}\tau
	\right.\nonumber\\
	&\left.+2L_{0}\alpha-\alpha t+2\tau\alpha+2L_{0}-t-2\tau\right].
\end{align}

The formula for $E^{\text{(cas)}}_{(\text{vac})}\left(t\right)$,
for $\beta\leq t<t_{b}^{(1)}$, is given by
\begin{equation}
	E^{\text{(cas)}}_{(\text{vac})}\left(t\right) =\begin{cases}
		E^{\text{(cas)}}_{\text{(DAC)}} & \left(\beta\leq t<t_{\text{enc}}\right),\\
		E^{\text{(cas)}}_{(\text{DC}1)}(t) & \left(t_{\text{enc}}\leq t<t_{N}\right),\\
		E^{\text{(cas)}}_{(\text{DC}2)}(t) & \left(t_{N}^{(1)}\leq t<t_{\text{des}}\right),\\
		E^{\text{(cas)}}_{(\text{DBC})} & \left(t_{\text{des}}\leq t<t_{a}^{(2)}\right),\\
		E^{\text{(cas)}}_{(\text{DB}1)}(t) & \left(t_{a}^{(2)}\leq t<t_{N}^{(2)}\right),\\
		E^{\text{(cas)}}_{(\text{DB}2)}(t) & \left(t_{N}^{(2)}\leq t<t_{a}^{\prime(2)}\right),\\
		E^{\text{(cas)}}_{(\text{DBD})} & \left(t_{a}^{\prime(2)}\leq t<t_{b}^{\prime(1)}\right),\\
		E^{\text{(cas)}}_{(\text{BD1})}(t) & \left(t_{b}^{\prime(1)}\leq t<t_{N}^{(2)}\right),\\
		E^{\text{(cas)}}_{(\text{BD2})}(t) & \left(t_{N}^{(2)}\leq t<t_{b}^{(1)}\right),
		\label{eq:E-out-vac-vac}
	\end{cases}
\end{equation}
where:
\begin{align}
	E^{\text{(cas)}}_{(\text{DAC})}=-\frac{\pi}{24}\frac{(\alpha\beta+L_0-2\tau-2\tau\alpha)}
	{(\alpha\beta+L_0)^2},
	\label{eq:E-cas-DAC}
\end{align}
\begin{align}
	E^{\text{(cas)}}_{(\text{DC1})}(t)=\frac{\pi}{24}\frac{(\alpha\tau+L_0+\beta-2t+\tau)}
	{(\alpha\beta+L_0)^2},
	\label{eq:E-cas-DC1}
\end{align}
\begin{align}
	E^{\text{(cas)}}_{(\text{DC2})}(t)=-\frac{\pi}{24}
	{\frac{\left(2\alpha\beta-\alpha\tau+3{L_0}
	+\beta-2t-\tau\right)}{\left(\alpha\beta+{L_0}\right)^{2}
	}},
	\label{eq:E-cas-DC2}
\end{align}
\begin{align}
	E^{\text{(cas)}}_{(\text{DBC})}=E^{\text{(cas)}}_{(\text{DAC})},
	\label{eq:E-cas-DBC}
\end{align}
\begin{align}
	E^{\text{(cas)}}_{(\text{DB1})}(t)=\frac{\pi}{24}
	{\frac{\left( 2\alpha\tau+{L_0}-t+2\tau
	\right)}{\left( \alpha\beta+{L_0} \right) ^{2}}},
	\label{eq:E-cas-DB1}
\end{align}
\begin{align}
	E^{\text{(cas)}}_{(\text{DB2})}(t)=-\frac{\pi}{24}
	{\frac{\left(2\alpha\beta-\alpha\tau+3{L_0}
	-t-\tau\right)}{\left(\alpha\beta+{L_0}\right)^{2}}},
	\label{eq:E-cas-DB2}
\end{align}
\begin{align}
	E^{\text{(cas)}}_{(\text{DBD})}=-\frac{\pi}{24}
	{\frac{\left(\alpha\beta-2\alpha\tau+{L_0}-2
	\tau\right)}{\left(\alpha\beta+{L_0}\right)^{2}}},
	\label{eq:E-cas-DBD}
\end{align}
\begin{align}
	E^{\text{(cas)}}_{(\text{BD1})}(t)=-\frac{\pi}{24}
	{\frac{\left(\alpha\tau+\beta-t+\tau \right)}{
	\left( \alpha\beta+{L_0}\right)^{2}}},
	\label{eq:E-cas-BD1}
\end{align}
\begin{align}
	E^{\text{(cas)}}_{(\text{BD2})}(t)=-\frac{\pi}{24}
	{\frac{\left(2\alpha\beta-2\alpha\tau+2{L_0}+
	\beta-t-2\tau\right)}{\left(\alpha\beta+{L_0}\right)^{2}}}.
	\label{eq:E-cas-BD2}
\end{align}

%
\section{Some details of the application to a concrete law of motion}
\label{ap:details-application}

Let us divide the regions in each Fig. \ref{fig:T-A-tilde-tau-10-1} and \ref{fig:T-B-tilde-tau-10-1}
into: vacuum regions I (vac-I) and II (vac-II), corresponding, respectively, to the left and right lighter regions in each one of these figures; in a similar way, particle regions I (par-I) and II (par-II), corresponding, respectively, to the left and right darker regions.
After integrating the energy densities in Fig. \ref{fig:T-A-tilde-B-tilde-10-1},
in each of their respective regions, we have:
$E_{\cal \tilde{A}\text{(vac-I)}}^{\text{(vac)}}\left(\beta\right)\approx -0.016$; 
$E_{\cal \tilde{A}\text{(vac-II)}}^{\text{(vac)}}\left(\beta\right)\approx -0.049$;
$E_{\cal \tilde{A}\text{(vac)}}^{\text{(vac)}}\left(\beta\right)=E_{\cal \tilde{A}\text{(vac-I)}}^{\text{(vac)}}\left(\beta\right)+E_{\cal \tilde{A}\text{(vac-II)}}^{\text{(vac)}}\left(\beta\right)\approx-0.065$;
$E_{\cal \tilde{A}\text{(par-I)}}^{\text{(vac)}}\left(\beta\right)\approx -0.002$;
$E_{\cal \tilde{A}\text{(par-II)}}^{\text{(vac)}}\left(\beta\right)\approx -0.012$;
$E_{\cal \tilde{A}\text{(par)}}^{\text{(vac)}}\left(\beta\right)=E_{\cal \tilde{A}\text{(par-I)}}^{\text{(vac)}}\left(\beta\right)+E_{\cal \tilde{A}\text{(par-II)}}^{\text{(vac)}}\left(\beta\right)\approx -0.014$;
$E_{\cal \tilde{B}\text{(par-I)}}\left(\beta\right)\approx 1.349$;
$E_{\cal \tilde{B}\text{(par-II)}}\left(\beta\right)\approx 1.349$;
$E_{\cal \tilde{B}}\left(\beta\right)=
E_{\cal \tilde{B}\text{(par-I)}}\left(\beta\right)+
E_{\cal \tilde{B}\text{(par-II)}}\left(\beta\right)\approx 2.699$;
$E^{\text{(cas)}}\left(\beta\right)\approx-0.054$.
Considering the length $l$ of each particle region $l=0.155$, we have $\overline{{\cal T}}_{\text{(par)}}^{\text{(vac)}}\left(t,x\right)
=(E_{\cal \tilde{A}\text{(par)}}^{\text{(vac)}}\left(\beta\right)+E_{\cal \tilde{B}\text{(par)}}^{\text{(vac)}})/(2l)\approx0.033$.


%

\end{document}